\documentclass[preprint,12pt,authoryear]{article}
\usepackage{ijcai21}

\usepackage{times}
\usepackage{soul}
\usepackage{url}
\usepackage[hidelinks]{hyperref}
\usepackage[utf8]{inputenc}
\usepackage[small]{caption}
\usepackage{graphicx}
\usepackage{amsmath}
\usepackage{amsthm}
\usepackage{booktabs}
\usepackage{algorithm}
\usepackage{algorithmic}
\title{Generative Transfer Learning: Covid-19 Classification with a few Chest X-ray Images}

\author{
	Suvarna Kadam$^1$
	\and
	Vinay G. Vaidya$^2$
	\affiliations
	$^1$$^2$Department of Technology, SPPU Pune INDIA \\

	\emails
	suvarna.kadam@gmail.com,
	vaidya.vinay@gmail.com
}

\begin{document}

\maketitle

\begin{abstract}
Detection of diseases through medical imaging is preferred due to its non-invasive nature. Medical imaging supports multiple modalities of data that enable a thorough and quick look inside a human body. However, interpreting imaging  data is often time-consuming and requires a great deal of human expertise. Deep learning models can expedite  interpretation and alleviate the work of human experts. However, these models are data-intensive and require significant labeled images for training. During novel disease outbreaks such as Covid-19, we often do not have the required labeled imaging data, especially at the start of the epidemic.  Deep Transfer Learning addresses this problem by using a pretrained model in the public domain, e.g. any variant of either VGGNet, ResNet, Inception, DenseNet, etc., as a feature learner to quickly adapt the target task from fewer samples. Most pretrained models are deep with complex architectures. They are trained with large multi-class datasets such as ImageNet, with significant human efforts in architecture design and  hyper parameters tuning.  \\
We presented \footnote{AI4SG-21: The 3rd Workshop on Artificial Intelligence for Social Good,  IJCAI 2021, Montreal Canada.\url{https://amulyayadav.github.io/AI4SG2021/}} a simpler generative source model,  pretrained on a single but related concept, can perform as effectively as existing larger pretrained models. We demonstrate the usefulness of generative transfer learning that requires less compute and training data, for Few Shot Learning (FSL)  with a Covid-19 binary classification use case. We compare classic deep transfer learning with our approach and also report FSL results with three settings of 84, 20, and 10 training samples. The model implementation of generative FSL for Covid-19 classification is available publicly at https://github.com/suvarnak/GenerativeFSLCovid.git.
\end{abstract}

\section{Introduction}
\label{sec1}
During Covid-19 pandemic, pneumonia has become part of differential diagnosis for patients  exhibiting respiratory distress along with other common symptoms \cite{Cleverleym2426}. Although many  people with Covid-19 infection may never develop pneumonia, predicting pneumonia is vital if present. Chest radiography can help to detect and expedite pneumonia diagnosis. However, a single feature on chest radiograph is not specific enough to diagnose. Often a combination of features, such as multi-focal opacity, linear opacity with the presence of primarily bilateral consolidation is commonly observed in Covid-19 pneumonia \cite{Cleverleym2426}. Chest Imaging can confirm the presence of these features, but an expert radiologist needs to identify it. Deep Learning based models can help automate chest image analysis to tackle the increased need for expert radiologists or wherever such expertise is not readily available. 
\subsection{ Deep Transfer Learning based Covid-19 Diagnosis}
Early detection is often the key for controlling the outbreak of any new disease, including Covid-19. Deep Learning based models can analyze and predict from medical images but require a large number of labeled images. Unfortunately, novel disease outbreaks pose a unique challenge as very few known samples are available at the start of the outbreak. Hence, we often do not have sufficient  images of the disease to robustly train deep models. 
Deep Transfer Learning is commonly used to remedy this problem by pretraining a model with  features that are borrowed from the source dataset. Multiple recent studies \cite{horryIeeeXfer,minaee2020deep,Apostolopoulos_2020,pham2020comprehensive, ahuja2021deep} discuss the results of applying  transfer learning to predict Covid-19 from medical imaging. \cite{minaee2020deep} and  \cite{Apostolopoulos_2020} used Chest X-ray Images,  \cite{pham2020comprehensive} and \cite{ahuja2021deep} used CT scans, whereas \cite{horryIeeeXfer} experimented on multiple modalities of images; X-ray, CT  and Ultrasound. All these approaches use pretrained models such as ResNet18, ResNet50, SqueezeNet, and Densenet-121  as source models for transfer learning. 
%These source models are Convolutional Neural Networks (CNN) trained on ImageNet dataset. 
\subsection{Large Pretrained Models for Few Shot Learning (FSL)}
Deep models that are pretrained on large open datasets such as ImageNet and  their learned weights are publicly available for transfer learning. However, using  these models as source models may pose following issues.
\begin{enumerate}
	\item Commonly used pretrained models in existing transfer learning approaches are  discriminative, i.e. models are trained with an objective to discriminate among a closed set of concepts or classes. Such discriminative classifier gives more importance to features that discriminate between classes and may not learn important concept features. The objective function of such model is focused on telling the concepts apart rather than learn the defining features of those concepts. While it may be a good strategy if the downstream task is to classify from a closed group, such model might miss out on learning certain features that are important to learn the representation of a concept but may not be required to discriminate it in that closed group.  Learning such features  is useful for better representation learning for the class, and necessary if the class needs to be later identified from a larger group of classes. If a model is being trained as a feature learner for other models, it is especially important that it learns a good representation with all features and not just the ones useful for discrimination. 
	\item Another issue with existing pretrained models is that they are often large and  trained at very high human effort, and computation costs. Although publicly available,  only entities with access to compute can actually train them. This goes against the principle of democratizing AI as pretrained model's design, training, and availability is invariably influenced by entities that can train it. Anyone using these models can have very little say over model's architecture, source data, and training process.
\end{enumerate}
In addition to these technical issues, use of pretrained models on public domain data is ethically ambiguous. It is unclear and not debated enough if such models should be used for commercial or proprietary purposes. Commercial use of public pretrained models can be ethically challenged by image copyright owners, as often model trainers do not own the data. For example, the ImageNet dataset description clearly states that it provides original images for non-commercial research \& educational use, and that it does not own the copyright of the images. Therefore, although these model weights are freely available at the moment, it may not  be the case in future. If pretrained models are made strictly open, then target models should  be made open too. This may pose a problem for the healthcare domain where model designers may be constrained to create target models with closed data for safeguarding patient privacy. In this paper, we demonstrate an alternate way to perform deep transfer learning with greater flexibility to choose model architecture and source data. We also show how a simpler generative source model that requires much less compute can be trained and used effectively as a custom pretrained model. 
\section{Generative FSL for Covid-19 Classification}
Generative Approach for Few-Shot Learning \cite{source_generativeFSL} attempts to mimic human-like \textit{intuitive} transfer learning by 1) Pretraining a single-concept generative models for class(es) that have sufficient labeled data for training, 2) Learning target concept that has fewer training samples, with the help of borrowed features from most related pretrained model. In this paper, we validate the generative FSL approach for predicting Covid-19  from chest X-ray images. Historically, unsupervised feature learning has been explored  for improving supervised task performance \cite{paine2014analysis,bengio2012unsupervised}. \cite{erhan2010does} experimented unsupervised feature learning with autoencoders to achieve better generalization from the training data. Generative FSL also advocates unsupervised feature learning. However, it proposes to pretrain a \textit{related generative model}.  As the features of the domain are prelearned with the generative model, the target task is classification can be learned even from \textit{fewer} samples. In our experimentation, we employed generative FSL for Covid-19 classification as a three-step approach, 1) Pretraining a suitable \textit{generative} source model 2) Creating target Covid-19 classifier with pretrained weights of generative model, and 3) Fine-tuning only classification layers of the Covid-19 classifier.
We compared our results with existing deep transfer learning based Covid-19 classifier.
\subsection{Pretraining Generative Source Model}
\begin{figure*}
	\includegraphics[width=\linewidth]{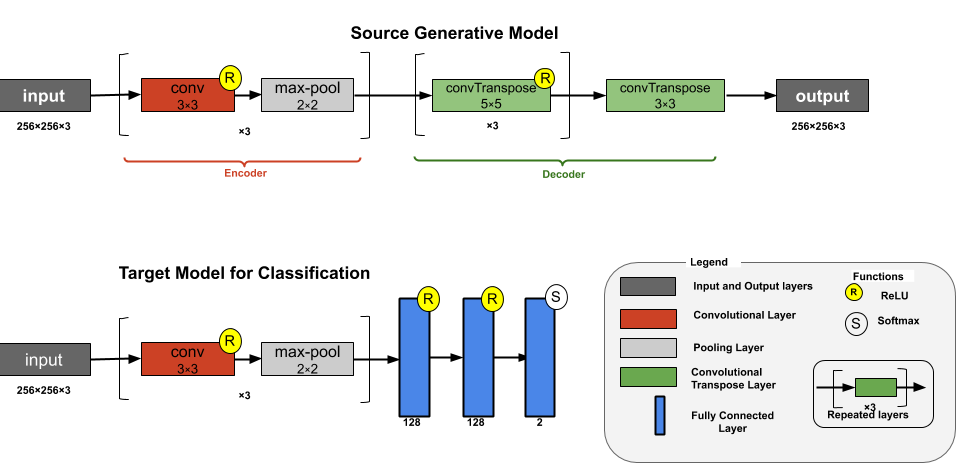}
	\caption{Generative FSL: Proposed Architecture for Source Pretrained Model \& Target Classifier}
	\label{proposedarch}
\end{figure*}

\begin{figure}[h!]
	\includegraphics[width=\linewidth]{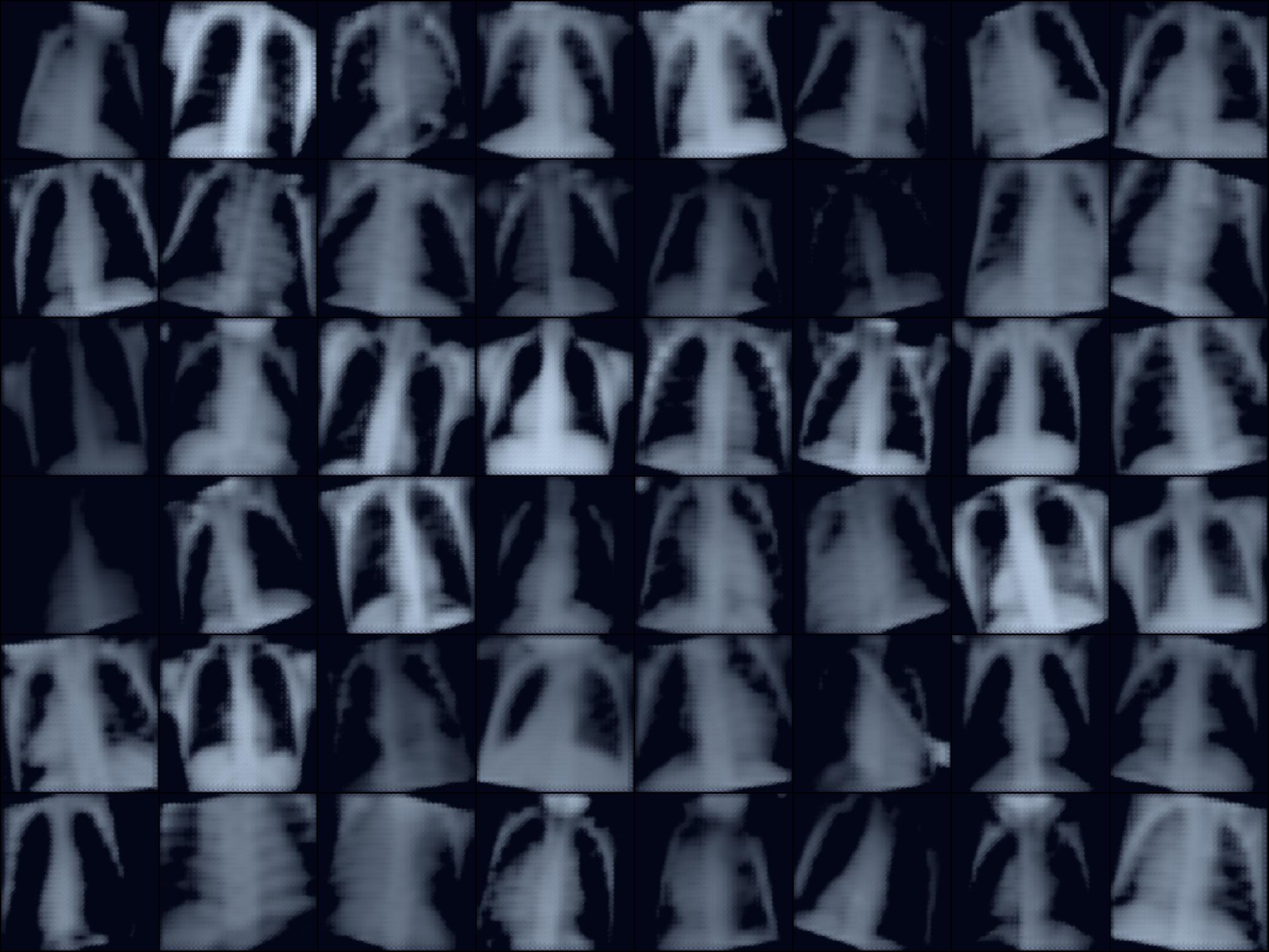}
	\caption{Sample Chest Images generated with model trained for 1 epoch}
	\label{fig:generated_0epoch}
\end{figure}

\begin{figure}[h!]
	\includegraphics[width=\linewidth]{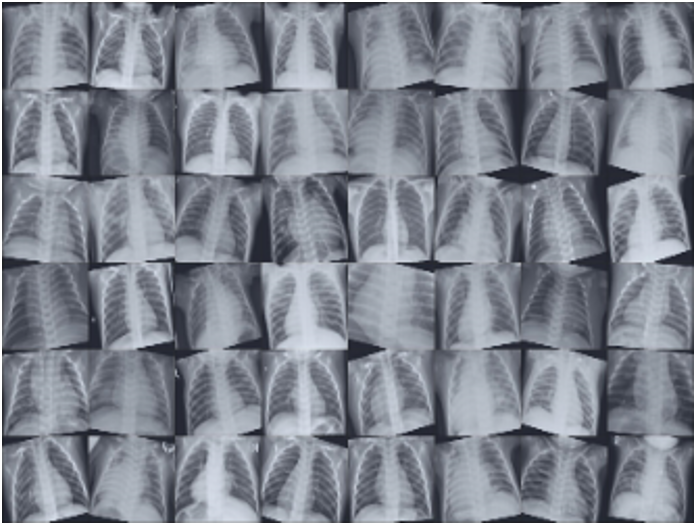}
	\caption{Sample Chest X-ray Images generated with model trained for 20 epoch}
	\label{fig:generated_20epoch}
\end{figure}

Almost all novel disease classification problems require FSL due to unavailability of sufficient training samples at the start of an outbreak. Therefore, Covid-19 binary classification is a good representative case of the FSL problem. We apply the generative FSL approach where, 1) We pretrain a single-concept generative model with ample available chest X-ray images. Figure \ref{proposedarch} illustrates proposed source and target  model architectures. The \textit{Source Generative Model } which we call ChestXrayGenNet is an architecturally simple \textit{Convolutional Autoencoder} \cite{kerasAE2016}.
Convolutional Autoencoders are generally trained to learn a compressed latent representation of image dataset such that \textit{any} sample image of that dataset can be regenerated. ChestXrayGenNet is trained with the Kaggle dataset of chest X-ray images \cite{chestxraykaggle}. Table \ref{tab:srctable} shows the exact number of images per split. 
\begin{table}[h!]
	\caption{Image Counts of Source Dataset: Chest X-ray  \protect \cite{chestxraykaggle} }
	\label{tab:srctable}
	\centering
	\begin{tabular}{|p{1cm}||p{1.25cm}|p{1.9cm}|}
		\hline
		\textbf{Split} & \textbf{Normal} & \textbf{Pneumonia}\\
		\hline
		Train & 1341 & 3875\\
		Test & 234 & 390\\
		\hline
	\end{tabular}
\end{table}
We use the source dataset to train our source generative model that has an objective to regenerate the chest X-ray images. It has an \textit{encoder} that learns to compress the input chest X-ray image into a smaller  representation and a \textit{decoder} that can re-generate image from that representation.  By forcing it to learn a smaller latent representation, it is ensured that only features that most effectively represent the concept of \textit{Chest X-ray} are learned.  Note that the chest X-ray dataset has images belonging to 2 classes, normal and pneumonia. However, we used its training images  to reconstruct, irrespective of image's class, and therefore we treated the dataset as a single-concept image collection for chest X-rays. Since autoencoders learn data-specific latent representation, they can generate data similar to what they have been trained on. In our case, the source model learns a good representation of the general concept of chest X-rays. Figure \ref{fig:generated_0epoch} shows a batch of generated test images after the model is trained with just 1 epoch. It shows that the model can generate images with identifiable features of chest X-ray, but has not yet learned the finer details . 
Figure \ref{fig:generated_20epoch} illustrates a batch of generated images after the model is trained for 20 epochs. We have used image transformations to augment the data such as rotating by $15^0$,  horizontal flipping and randomized cropping while training and validating the model. 
During source model training, Generative FSL differs from classical deep transfer learning in two aspects; 1) We can train the source model without any specialized hardware acceleration, e.g. ChestXrayGenNet can be trained on a regular laptop. 2) Our source model is generative and trained to have minimum reconstruction loss.
\subsection{Creating Target Covid-19 Classifier}
After pretraining, the learned weights of the `Encoder' part of the source model is used to create and initialize target classifier. Figure \ref{proposedarch}  illustrates the target model's architecture, which is exactly same as the \textit{encoder} part of the source model, with two additional fully connected (FC) linear layers. We trained  target classifier model for Covid-19 prediction by fine-tuning only the  last two layers. We are using the dataset by \cite{minaee2020deep}, COVID-Xray-5k Dataset which was compiled from two datasets and made publicly available. This dataset has chest X-ray images belonging to 2 classes, Covid  and non-Covid. Table \ref{tab:targettable} shows the exact number of images per split. 
\begin{table}[h!]
	\caption{Target Dataset Description COVID-Xray-5k  \protect \cite{minaee2020deep}}
	\label{tab:targettable}
	\centering
	\begin{tabular}{|p{1cm}||p{1.25cm}|p{1.9cm}|}
		\hline
		\textbf{Split} & \textbf{Covid} & \textbf{Non-Covid}	\\		\hline
		Train & 84 & 2000\\ 
		Test & 100 & 3000\\
		\hline
	\end{tabular}
\end{table}
%\subsection{Inherent Class Imbalance in Novel Disease Imaging Dataset }
As evident from the table, the target dataset has a severe class imbalance as it has less images with positive diagnosis of Covid as compared to non-Covid.  In fact, most imaging datasets for novel disease have highly skewed class distribution with inherent class imbalance.  Such imbalanced datasets pose a unique challenge while training deep models that otherwise can achieve good predictive performance. Most machine learning including deep models have an implicit assumption that the dataset is balanced. If not, models perform poorly, specifically for the minority classes (i.e. classes with fewer samples). This is a significant issue as often we are interested in the minority class, e.g. in Covid-19 prediction, we are more interested in correct predictions of \textit{Covid-19 positive} case. Class imbalance is a known issue in machine learning \cite{krawczyk2016learning, Chawla_2002}. Deep Learning-based classification methods employ coping strategies such as class re-sampling or threshold moving to improve model performance. \cite{Chawla_2002} proposed Synthetic Minority Over-sampling Technique  (SMOTE) where either under-sampling of the majority (normal) class or oversampling of the minority class is advocated to improve sensitivity to the minority class.  Another technique to tackle class imbalance is threshold moving \cite{COLLELL2018330}. They argued that re-sampling methods can cause asymmetric changes to samples and thus can introduce their own biases. An alternative suggested was threshold moving technique, which advocates adjusting the threshold that separates classes. \cite{krawczyk2016learning} categorized the learning approaches from imbalanced data into three categories; 1) Data-level methods that attempts to balance the samples' distributions, e.g. the oversampling of minority class,  2) Algorithmic-level methods that attempt to reduce the learner's bias towards majority class so that it can adapt to skewed data, e.g. Threshold moving or cost sensitive learning where learner pay varying penalty for (not) learning each class. and 3) Hybrid methods combine both approaches, e.g.  both sampling and cost-sensitive learning applied to improve performance. In our experimentation, we employed SMOTE by implementing weighted sampling of training data as it is the simplest method to handle imbalance. We computed the imbalance factor as a ratio between image counts of Covid and Non-Covid classes and used it to balance skewed Covid-19 X-ray imaging data during the training. 
\subsection{ Generative Transfer Learning for Training  Classifier}
\begin{figure}[h!]
	\includegraphics[width=\linewidth]{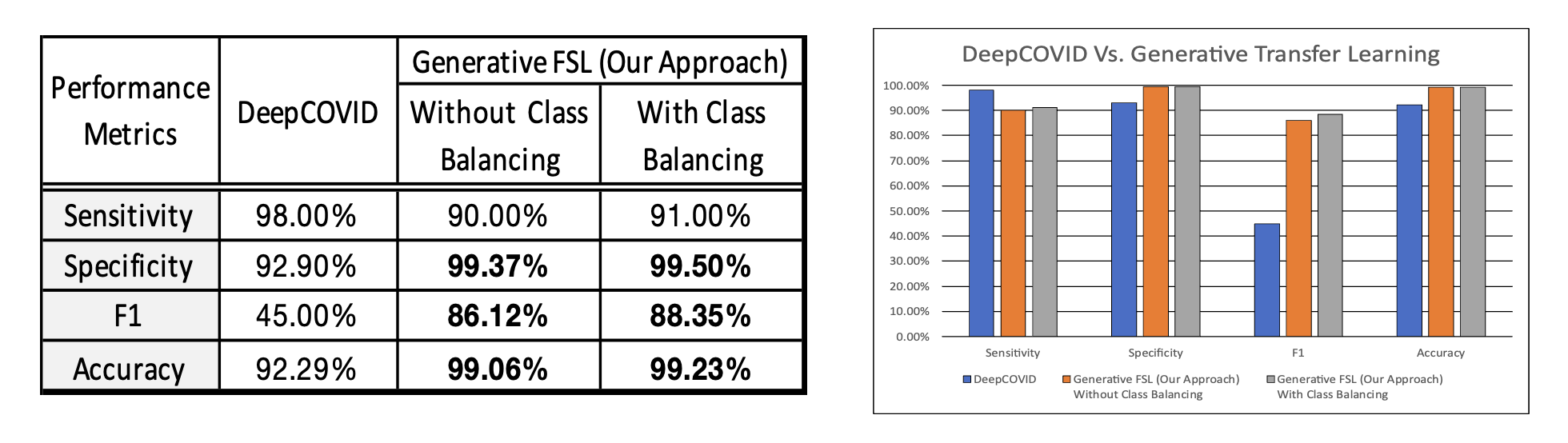}   %genetrativeTL.png}
	\caption{Comparing Performance of Covid-19 classifier with Discriminative Vs. Generative Pretrained Source Models}
	\label{fig:generativeTL}
\end{figure}

We trained the target classier similar to classic deep transfer learning, 1) We initialized the classifier with learned weights of the encoder of ChestXrayGenNet and 2) We trained only the last two fully connected layers (Blue layers in Figure \ref{proposedarch})  with target dataset COVID-Xray-5k \cite{minaee2020deep}. This dataset was compiled from two datasets and curated with human radiologists to  include, 2,084 training (84 Covid, 2000 Non-Covid) and 3,100 test images (100 Covid, 3000 Non-Covid).  Effectiveness of deep transfer depends on selection of relevant pretrained model, e.g. a model trained on the source visual concepts that are similar to the target concept. \\
Figure \ref{fig:generativeTL} shows the comparison of classic transfer learning and our approach of generative transfer learning, across performance parameters for Covid-19 classification. We compared the results with DeepCovid \cite{minaee2020deep} as it uses a discriminative pretrained model. 
As the Figure \ref{fig:generativeTL} illustrates that we  got comparable results with a much simpler model. Our approach yields better performance for Specificity, F1 and Accuracy.   
\subsection{Generative Transfer Learning to Tune  Classifier with Fewer Shots}
We observed  that a simpler  generative source model \cite{source_generativeFSL} performs comparatively the same or better than classic ImageNet based larger pretrained models for Covid-19 classification. Extending the observation further, we hypothesized that since we use  a source generative model that is trained on large and related dataset of Chest X-ray images \cite{chestxraykaggle}, it may enable the target classifier to learn with \textit{fewer} samples.  To validate the hypothesis, we  finetuned Covid-19 target classifier by using fewer training samples. Specifically, we trained the target classifier in the following 5 different settings.
\begin{enumerate}
	\item All-shots:  All available training samples are used to finetune the FC layers of target classifier. 
	\item 84-shots:  Maximum equal number of Covid and Non-Covid  training samples are used to finetune  FC layers of classifier. Since the dataset has a maximum of 84 samples for the minority Covid class, we randomly chose 84 non-Covid samples for finetuning.
	\item 20-shots:  Randomly selected 20 samples of each class are used for finetuning. 
	\item 10-shots:  Randomly selected 10 samples of each class are used for finetuning. 
	\item All-shots with balanced data loading:  Randomly selected, equal number of samples are selected in every batch during the training of target classifier.
\end{enumerate}
We compared our generative FSL approach with classical Deep transfer learning for Covid Classification. Table \ref{tab:compareresults} illustrates performance metrics, sensitivity and specificity along with F1 score and accuracy.
\begin{table*}[h!]
	\caption{Deep Transfer Learning for Covid-19 Classification:  Generative FSL Vs. Deep-COVID \protect \cite{minaee2020deep} }
	\label{tab:compareresults}
	\centering                         
	\includegraphics[width=\linewidth, height=1.95in]{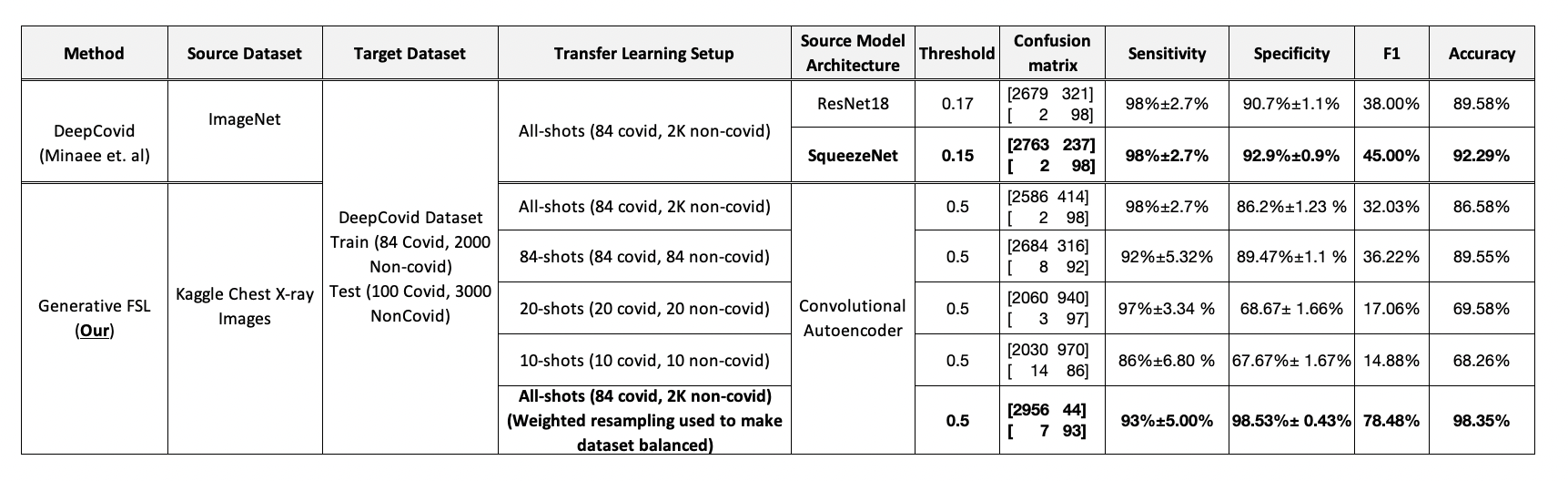}
\end{table*}
\section{Discussion of  Results}
\begin{figure}[h!]
	\includegraphics[width=\linewidth]{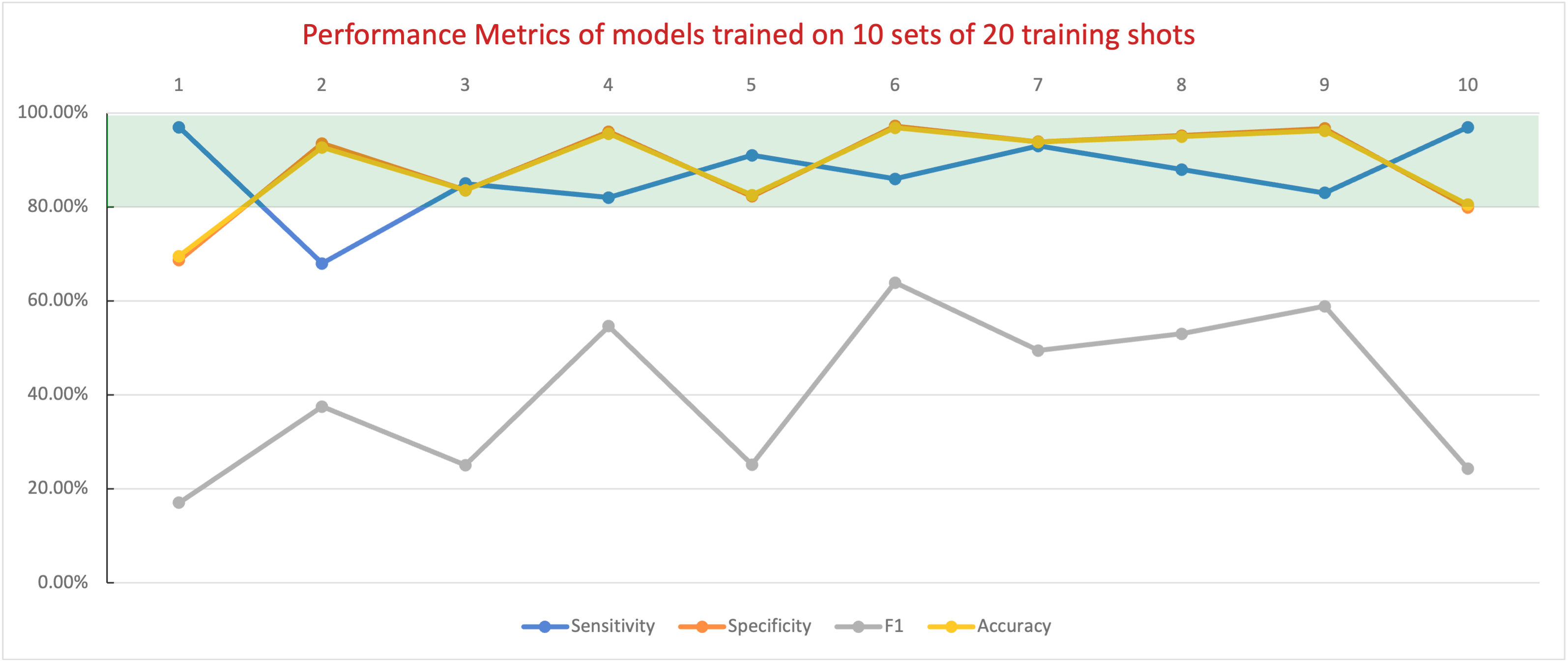}
	\caption{Variation in Performance Metrics of models trained with random 20 training images per class}
	\label{fig:chart_random20_xray}
\end{figure}

\begin{figure}[h!]
	\begin{center}
		\includegraphics[width=8cm]{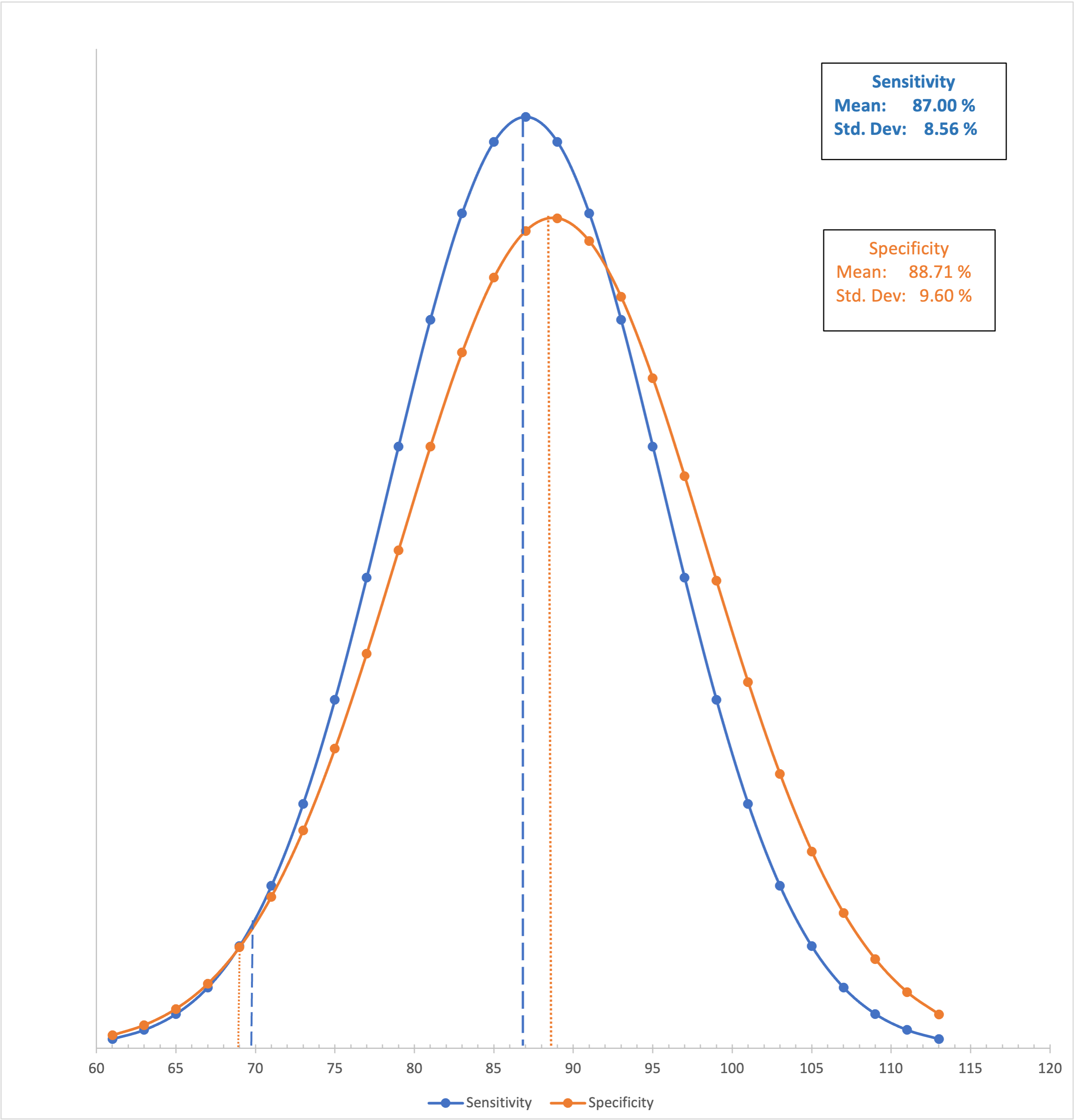}
		\caption{Sensitivity \& Specificity Distributions for 10 Models Trained on Random Subsets of 20-shots}
		\label{fig:ssdistribution}
	\end{center}
\end{figure}

Table \ref{tab:compareresults} illustrates that we get comparable results with our approach when all training samples are used.  We observed that without much hand-engineering for hyper parameters, the model out-performed in terms of specificity, F1 score and accuracy. 
Since the target dataset has much fewer Covid-19 test images and therefore performance measurements may not be reliable as mentioned in the reference work of Deep-Covid \cite{minaee2020deep}. Still  the results demonstrate that generative FSL approach can provide novel disease classification with much better trade-off between Sensitivity and Specificity. We also computed the metrics by moving the threshold, but we observed best performance at 0.5 as we used weighted sampling. Generative FSL yields encouraging results for Covid-19 classification in the  extreme scenarios of just 20 or 10 training images. Our approach is much simpler when compared to classic deep transfer learning including  Deep-Covid  as it does not require,  1) Extensive compute, hand engineering while training source models, 2) Analysis of multiple ImageNet based models on the target task, or 3)Threshold-moving to see which value provides the best Sensitivity. Our approach simplifies the training pipeline, and therefore makes it easy to automate and scale FSL for \textit{any} novel class prediction use case. \\
In this experimentation, we have not employed domain-specific human expertise to improve performance as our efforts were focused on verifying whether generative FSL can be applied for Covid-19 predictions. But models trained by our approach can be further improved with better data.  As current understanding of Covid-19 is fluid, with no fixed definition of Covid-19 induced pneumonia \cite{Cleverleym2426}, detection  based only on chest X-ray are not recommended. Moreover, Interpreting Covid-19 induced pneumonia from X-ray images can be ambiguous and challenging even for an expert radiologist due to closely related patterns displayed in Chest X-ray of pneumonia caused by Covid or other conditions. However, pandemic still continues and Covid-19 is likely to remain an important differential diagnosis for people presenting  with other known symptoms of the disease. Therefore, quick and automated diagnosis based on Chest X-ray is desirable even with concerns about the reliability of results. Moreover reliability of the results can be improved with more and better image data. Another reason for unreliability of results is that the test split of target dataset COVID-Xray-5k  is also heavily imbalanced. For example, we are computing Sensitivity for just 100 Covid images and Specificity with 3000 non-Covid images. We computed confidence intervals for more clarity on results and observed wider intervals for models trained on fewer samples. This is expected as selecting fewer training samples introduces more randomness due to different ways we can select the samples. To ensure that Performance of models trained on 20-shots and 10-shots is not due to cherry-picking better image data, we conducted another experiment where we created 10 random data-subsets of 20 shots of Covid and non-Covid images. We trained 10 separate models on each of these subsets and computed the performance metrics. Table \ref{fig:random20_xray} illustrates the metrics of these 10 models.
\begin{table}[h!]
	\caption{Performance of 10 Models Trained on Random Subsets of 20-shots}
	\label{fig:random20_xray}
	\centering
	\includegraphics[width=9cm]{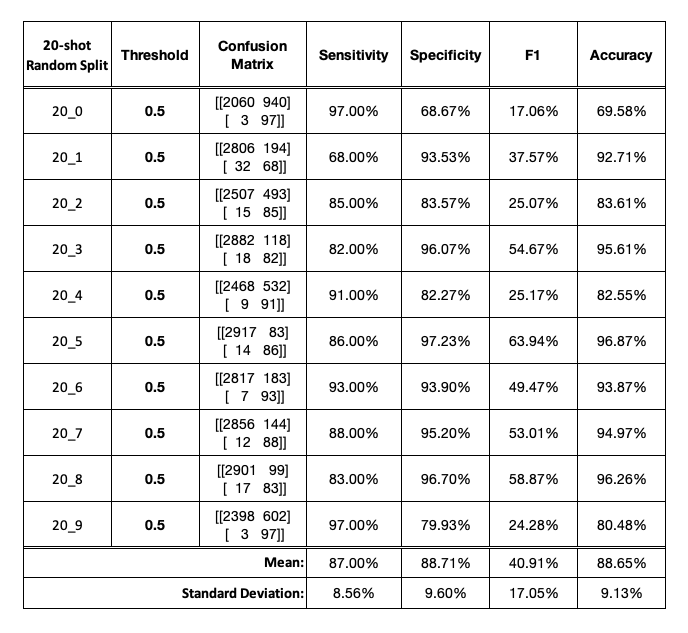}
\end{table}
To further assess the reliability of our 20-Shot results, we analyzed the variation in Sensitivity, Specificity, F1 and Accuracy. Figure \ref{fig:chart_random20_xray} illustrates the variation across all 10 models. 
While F1 score varied widely, Sensitivity, Specificity and accuracy mostly remained above 80\% (green band in Figure \ref{fig:chart_random20_xray}) for 20-shot models trained with the generative FSL approach. Another interesting conclusion is that accuracy is strongly dominated by specificity and almost mimics it. This mainly due to the significant imbalance in test data we used.  We have significantly more non-Covid (3000) samples compared to Covid (100) samples. Therefore, performance in the positive class (Covid) has a low influence on accuracy and F1 score. In fact, F1 almost mimics pattern of specificity. Figure \ref{fig:chart_random20_xray} also highlights that we need to carefully analyze the sensitivity and F1 measured on imbalanced datasets such as COVID-Xray-5k . Our analysis also confirms the claim
by  \cite{Chicco2020}  that although accuracy and F1 score are the  most adopted metrics in binary classification tasks, these can be misleading and give inflated results especially on imbalanced datasets. Results on imbalanced datasets should be reported either by balanced loading of test images while computing the performance metrics or with robust metrics such as  Matthew’s Correlation Coefficient (MCC) for fair comparison and benchmarking. We plotted the distribution of sensitivity and specificity of 10 models to get a visual understanding of the variation. Figure \ref{fig:ssdistribution} illustrates the  distribution of both metrics with mean and standard deviation.
This figure visually shows that variation is much wider than common practice of computing confidence interval as shown in Table \ref{tab:compareresults} and it helps us to make a more reasonable claim that the generative FSL approach for 20-shot Covid positive prediction is about 87\%±8.56\%. 
\section{Conclusion}
We reported a generative Few Shot Learning approach for Covid-19 detection from Chest X-ray images by pretraining Convolutional Autoencoder model on Kaggle Chest X-ray Images \cite{chestxraykaggle}. We created a Covid 19 classifier with pretrained layers of custom source model, stacked with  added classification layers and finetuned it with publicly available COVID-Xray-5k dataset. We evaluated the finetuned model and reported the comparative results with a Deep Transfer Learning based Covid-19 binary classifier. We  evaluated the classifier with multiple  performance metrics; Sensitivity, Specificity and  F1 score, and reported results by  training it with 4 different setups that use either all, 84, 20 or 10 training samples for learning. \\
Though our work is evaluated on dataset with only 100 Covid-19 positive images, our results are encouraging as the approach shows promise of using a simpler source  model  for Few Shot classification task. The generative FSL approach can be applied to \textit{any} FSL Image classification problem. We chose to demonstrate it on Covid-19 classification from  X-ray images due to the current pressing need to solve it. Our study also demonstrated the possibility of applying deep learning to solve a challenging problem in a low compute and low training-data setup.  Due to the limited number of Covid-19 images publicly available so far, further experimentation is needed on the larger dataset for more reliable evaluation of model performance.\\
Our work can be extended in several ways to improve Covid-19 prediction from few training samples. Similar to chest X-ray images, source generative model can be trained with CT images and used as domain feature learner to pretrain the target classifier. Our work can be  used to benchmark FSL with more robust metrics such as Matthew’s Correlation Coefficient.
% for bibtex

\bibliographystyle{named}
\bibliography{refs1}


\end{document}